\documentclass[preprint, aps, floats, preprintnumbers, groupedaddress, superscriptaddress,floatfix, nofootinbib]{revtex4}
\usepackage{amsmath,amssymb,epsfig,color,bbold,physics,slashed}
\usepackage{mathrsfs}  
\usepackage[capitalize]{cleveref}

\def\iu{{\rm i}} 
\def\e{{\rm e}}

\usepackage[force]{feynmp-auto}
\usepackage{psfrag}
\usepackage{graphicx}
\usepackage{bm}

\newcommand{\be}{\begin{equation}}  
\newcommand{\ee}{\end{equation}}

\allowdisplaybreaks

\begin{document}

\begin{titlepage}

\begin{flushright}
\end{flushright}

\preprint{CALT-TH-2024-019
}

\vspace{0.7cm}
\begin{center}
\Large\bf 
The effective field theory of extended Wilson lines 
\end{center}

\vspace{0.8cm}
\begin{center}
{\sc   Ryan Plestid }\\
\vspace{0.4cm}

{\it Walter Burke Institute for Theoretical Physics,\\
California Institute of Technology,
Pasadena, CA, 91125 USA\vspace{1.2mm}}
\end{center}
\vspace{1.0cm}

\author{Ryan Plestid}
\begin{center}
\today
\end{center}

\vspace{1.0cm}

  \vspace{0.2cm}
  \noindent
    We construct the effective theory of electrically charged, spatially extended, infinitely heavy objects at leading power. The theory may be viewed as a generalization of NRQED for particles with a finite charge distribution where the charge radius and higher moments of the charge distribution are counted as $O(1)$ rather than $O(1/M)$. We show this is equivalent to a Wilson line traced by the worldline of an extended charge distribution.  Our canonical use case is atomic nuclei with large charge $Z\gg 1$. The theory allows for the insertion of external operators and is sufficiently general to allow a treatment of both electromagnetic and  weak mediated lepton-nucleus scattering including charged-current processes.  This provides a first step towards the factorization of Coulomb regions, including structure dependence arising from a finite charge distribution, for scattering with nuclei.   

\end{titlepage}

\tableofcontents
\pagebreak

\section{Introduction}

The soft limit of gauge theories is fully characterized by an effective field theory (EFT) of Wilson lines, where particles are treated as infinitely heavy point-like 
 sources \cite{Caswell:1985ui,Korchemsky:1991zp,Isgur:1989vq,Isgur:1990yhj,Georgi:1990um,Falk:1990yz,Bauer:2002nz,Becher:2009qa,Grozin:2022umo}. Feynman rules involve eikonal propagators, and  allow one to dramatically simplify the analysis of the soft sector of a theory. This simplification underlies our ability to control radiative corrections for hard scattering processes, and  the shuffling of soft dynamics into Wilson lines is the basis of many factorization theorems \cite{Bauer:2002aj,Beneke:2003pa,Bauer:2004tj,Becher:2006mr,Becher:2008cf,Stewart:2009yx,Chiu:2012ir,Feige:2014wja,Rothstein:2016bsq}. 

When particles are very heavy, this is an especially powerful tool. For example, consider the scattering of $E\sim 1~{\rm GeV}$ charged particles off of a $M\sim 200~{\rm GeV}$ particle of charge $Z \gg 1$. The large charge of the heavy particle, $Z\gg1$, ensures that the  matrix element is dominated by ladder graphs and  since $E \ll M$ the heavy particle is effectively static. One can then make use of eikonal identities to reduce the leading series in $Z\alpha$ to a potential scattering problem \cite{Brodsky:SLAC1010,Dittrich:1970vv,Neghabian:1983vm,Weinberg:1995mt} dramatically simplifying the analysis. 

In the example above the particles we have in mind are heavy nuclei, e.g.\ lead, tungsten, gold etc. or any other heavy nucleus relevant for particle physics experiments. Unlike elementary particles, however,  heavy nuclei become larger (i.e.\ more spatially extended) as their mass increases. The scaling of length scales for a nuclei and point-like particles is given by 
\begin{equation}
    R\sim M^{1/3} \qq{(nucleus)} , \qq{vs.}~~ R\sim 1/M \qq{(point-like).} 
\end{equation}
Clearly the scaling of the radii are qualitatively opposite. The consequence of this scaling is that for heavy nuclei the inverse radius is typically $R^{-1} \sim 25~{\rm MeV}$ while their mass is $M\sim 200~{\rm GeV}$. The point-like limit then only applies for $Q \ll R^{-1} \ll M$, severely restricting the phase space where a point-like treatment is appropriate as opposed to $Q \ll M$ as in e.g., the case of a proton. In potential scattering models, one simply replaces the Coulomb field by a static charge distribution, solves Gauss' law, and  computes scattering in the potential. This allows one to extrapolate the static field approximation to larger values of $Q$, however it is not obvious how to interpret such a procedure in an EFT framework, or how to systematically improve the calculation to account for higher order radiative corrections.

\begin{figure}
    \centering
    \includegraphics[width=0.5\linewidth]{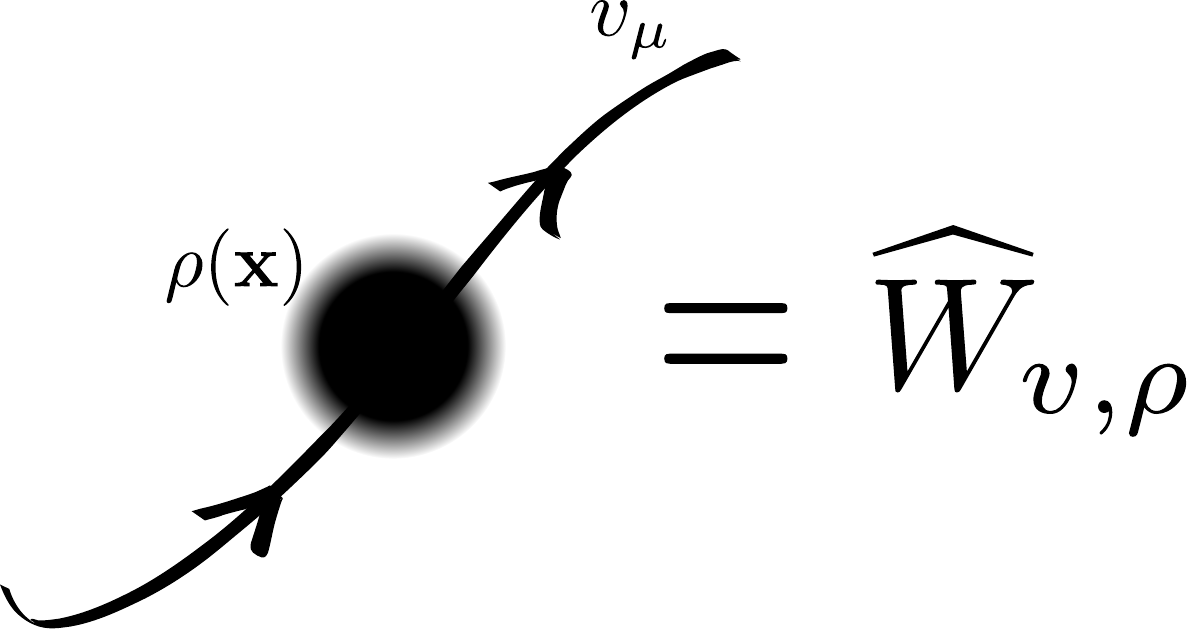}
    \caption{Schematic of extended Wilson line operator for an extended charge distribution $\rho(\vb{x})$ tracing out a trajectory along a worldline parameterized by the time-like vector $v_\mu$. This is a generalization of a typical point-like Wilson line. \Cref{ex-WL-def,chi-WL-def} define the appropriate non-local operator in general gauge. The gauge invariant Feynman rule is given by \cref{resum-feynmanrule}. \label{extended-WL-schematic} }
\end{figure}

The purpose of the present article is to construct an EFT applicable when $q R \sim O(1)$ while $q/M \ll 1$. This is the generalization of point-like Wilson lines to a theory of Wilson lines with extended structure as shown schematically in \cref{extended-WL-schematic}. We focus on QED because of its simplicity and relevance for applications to nuclear scattering and  bound-state problems; there are few, if any,  composite states more ubiquitous than atomic nuclei. We will restrict our discussion to the insertion of a single external operator.

We are motivated by experiments that require percent-level (or better) precision for reactions involving nuclei; in particular the neutrino oscillation program \cite{Branca:2021vis,Tomalak:2022xup} and super allowed beta decay \cite{Seng:2018yzq,Czarnecki:2019mwq,Hardy:2020qwl}. Other applications use nuclear targets and  explore extremely specialized pockets of phase space where radiative corrections may be larger than expected and  poorly understood; we have in mind Mu2e/COMET and  LDMX \cite{LDMX:2018cma,Bernstein:2019fyh}.   In these systems, long and  intermediate wavelength photons will experience an enhancement, proportional to $Z$, due to their coherent interaction with the nucleus.  When considering virtual QED corrections to high energy scattering with nuclei one must integrate over all scales ranging from the coherent to the incoherent regime, and  one therefore always probes the intermediate regime $q R \sim O(1)$.  

Precision measurements involving nuclei must account for the $Z$-enhanced contributions discussed above. To do so one must account for the composite's structure beyond tree level, i.e.\ inside of loops and  this necessitates a QFT formalism that explicitly includes the scale of nuclear structure inside higher order loop diagrams. When $1/R$ is well separated from other energy scales in the problem, then an EFT treatment can be formalized which separates coherently enhanced regions from high energy regions where $Z$-enhancements are absent and  conventional QED power counting holds. In what follows we study how structure dependence from large composite particles enters the loop expansion.  While nuclei are our main focus, our results apply to any theory with composite particles coupled to a $U(1)$ gauge theory. 

The rest of the paper is organized as follows. In \cref{Motivation} we sketch how coherently enhanced regions appear when considering QED loops in the ``full theory'' using electron-nucleus scattering as an example. We apply the method of regions and find that coherent enhancements occur in a particular momentum-region. In \cref{NRQED} we construct an EFT which allows for composite structure, and which reduces to NRQED in the point-like limit. We apply these ideas to $(e,e')$ scattering in the high energy limit in \cref{e-e'}. Finally in \cref{conclusions} we summarize our findings, and  suggest future directions of interest from both a formal EFT and  phenomenological perspective.

%

\section{Motivation and  setup \label{Motivation} }
Let us consider a reaction
\begin{equation}
    \label{X-reaction}
    A ~+~ X_1 \quad \rightarrow \quad B~+~X_2
\end{equation}
where $A$ and  $B$ are heavy composite objects, while $X_1$ and  $X_2$ represent probe states. Working in Feynman gauge, a generic ladder diagram\footnote{We focus on ladder diagrams here to illustrate the appearance of coherent enhancements. In the EFT we develop there is no restriction to a ladder topology. } in which $n$ photons are exchanged with the nucleus will be given by (with $\dd q_i = \dd^4 q/(2\pi)^4$) 
\begin{equation}
    \label{general-blobs}
    \begin{split}
    \iu\mathcal{M}_a^{(n)} = \int &\qty[ \dd q_i~ \frac{-\iu}{q_i^2}]  X^{\mu_1...\mu_n}_a (q_1, ..., q_n) \\
    &\int\qty[ \dd^4x_j ]  \e^{-\iu q_i \cdot x_i}  \mel{B(\vb{p}')}{T\{ J_{\mu_1}(x_1)... \mathcal{J}(0)...J_{\mu_n}(x_n)\}}{A(\vb{p})}~,
    \end{split}
\end{equation}
where $\mathcal{J}$ is an external current, $X^{\mu_1... \mu_n}_a$ represents some fixed set of photon insertions among the legs of the charged particles that make up either $X_1$ or $X_2$ in \cref{X-reaction}. The square brackets are a short hand for a product over indices.

\begin{figure}
    \centering
    \includegraphics[width=0.6\linewidth]{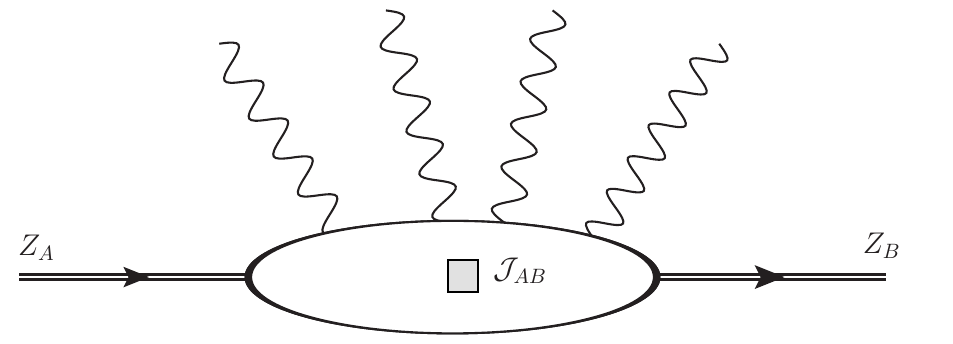}
    \caption{Diagram corresponding to \cref{correlator-def}. The gray box denotes the insertion of a single external current, and the blob includes all possible permutations of the photon vertices.  \label{blob-fig}}
\end{figure}

The time ordering  produces all possible crossed and uncrossed ladder diagrams. For QED applications, it suffices to study the correlator (shown diagramatically in \cref{blob-fig})
\begin{equation}
    \label{correlator-def}
    G_{\{\mu_i\}}(\{q_j\}) =\int\qty[ \dd^4x_j ~  \e^{-\iu q_j \cdot x_j}  ]\mel{B(\vb{p}')}{T\{ J_{\mu_1}(x_1)... \mathcal{J}(0)...J_{\mu_n}(x_n)\}}{A(\vb{p})} ~.
\end{equation}
For small photon momenta photon couplings can be enhanced by $Z_A$ and/or $Z_B$. We analyze the problem in the limit of $q/M \rightarrow 0$, but with $q R\sim O(1)$ held fixed. Our goal will be to identify contributions enhanced by factors of $Z$. 

Let us explicitly expand the time orderings by labeling the time orderings sequentially by $\{a,b,...\beta,\alpha\}$  such that $x_a<x_b<...<0<...<x_\beta<x_\alpha$. Here, Latin letters correspond to times before the hard current $\mathcal{J}(0)$ and Greek letters to times after the hard current. Next, insert a complete set of states between $J_{\mu_a}(x_a)$ and  $J_{\mu_b}(x_b)$. Integration over $\dd^3x$ enforces momentum conservation via a delta function, while the time integral gives
\begin{equation}
     ~\int_{-\infty}^{t_{b}} \dd t_a~ \e^{-\iu \Delta E_a t_a } \mel{A(\vb{p}+\vb{q}_a)}{J_{\mu_a}}{A(\vb{p})} = \frac{\e^{-\iu \Delta E_at_b}}{-\Delta E_a + \iu 0} \mel{A(\vb{p}+\vb{q}_a)}{J_{\mu_a}}{A(\vb{p})}~.
\end{equation}
where $\Delta E_a= \epsilon_A(\vb{p}+\vb{q}_a)-[\omega_a+\epsilon_A(\vb{p})]$ where $\epsilon_A(\vb{p})=(\vb{p}^2+M_A^2)^{1/2}$ is the on-shell energy of the heavy particle $A$. If we focus on the last photon in the time ordering we instead find
\begin{equation}
  ~\int^{\infty}_{t_{\beta}} \dd t_\alpha~ \e^{\iu \Delta E_\alpha t_\alpha } \mel{B(\vb{p})}{J_{\mu_\alpha}}{A(\vb{p}-\vb{q}_\alpha)}
   = \frac{\e^{\iu \Delta E_\alpha t_\beta}}{\Delta E_\alpha + \iu 0} \mel{B(\vb{p})}{J_{\mu_\alpha}}{A(\vb{p}-\vb{q}_\alpha)}~.
\end{equation}
where $\Delta E_\alpha =  \epsilon_B(\vb{p}+\vb{q}_\alpha)-[\omega_\alpha+\epsilon_B(\vb{p})]$. 
Carrying this procedure out term by term in each time ordering then yields analogous terms with e.g.\   $\Delta E_b =\epsilon_A(\vb{p}+\vb{q}_a+\vb{q}_b)-[\omega_a+\omega_b+\epsilon_A(\vb{p})]$ and  higher terms defined iteratively. After integrating over all times, the result is that\footnote{Note that \cref{exact-G} can be derived directly using old-fashioned perturbation theory.}
\begin{equation}
    \label{exact-G}
    \begin{split}
      G_{\{\mu_i\}}(\{q_j\}) =\sum_{\rm perms}\prod_{a}^{N_L} \frac{J_{\mu_a}^{[A]}(\vb{q}_{a})}{-\Delta E_a+ \iu 0} \prod_{\alpha}^{N_R}
      \frac{J_{\mu_\alpha}^{[B]}(\vb{q}_{\alpha})}{\Delta E_\alpha + \iu 0} ~\times \mathcal{J}_{AB}&
      \qty([\vb{p} +\Sigma \vb{q}_a]-[\vb{p}'-\Sigma \vb{q}_\alpha])~,\\
     &\hspace{0.1\linewidth}+ ({\rm other~ states})
     \end{split}
\end{equation}
where $J^{[A]}_{\mu_a}(\vb{q}_a)= \mel{A(\vb{p}+\vb{q}_a)}{J_{\mu_a}}{A(\vb{p})}$ and    $J^{[B]}_{\mu_a}(\vb{q}_a)= \mel{B(\vb{p}+\vb{q}_\alpha)}{J_{\mu_\alpha}}{B(\vb{p})}$ are on-shell matrix elements of the current $J_\mu$, and  ``other states'' refers to all intermediate states other than elastic scattering.\footnote{These states are subdominant to the elastic scattering transitions for highly charged objects, $Z\gg 1$.} The transition matrix element between $A$ and  $B$ depends, in general, on the momenta transferred via virtual photons,  
\begin{equation}
    \label{exact-JAB}
    \mathcal{J}_{AB}=\mel{B(\vb{p}'-\Sigma \vb{q}_\alpha)}{\mathcal{J}}{A(\vb{p}+\Sigma \vb{q}_a)} ~.
\end{equation}
The labels $N_L$ and  $N_R$ in \cref{exact-G} denote the number of photons inserted prior to (left of) and  after (right of) the hard current in a given time ordering. 

The most general matrix element (i.e.\ for arbitrary spin) of the electromagnetic current contains a unique form factor in the $M\rightarrow \infty$ limit \cite{Lorce:2009br,Lorce:2009bs}, 
\begin{equation}
    \mel{A(\vb{p}+\vb{q})}{J_{\mu}}{A(\vb{p})}= Z v_\mu  F(\vb{q}^2)  + O(q/M)~,
    ~\label{static-form-factors}
\end{equation}
where we have tacitly assumed that all higher order form factors that appear in  \cite{Lorce:2009bs} are normalized with $O(1)$ coefficients such that $(Ze) (|\vb{q}|^n/M^n)$ provides a reasonable estimate of their parametric size\footnote{E.g.\ the magnetic moment of nuclei is  of order $(Ze)/M \sim e/2m_N$ i.e.\ the nuclear magneton. } \cite{Lorce:2009br,Lorce:2009bs}.  Neglecting all recoil effects such that the $\Delta E_a \rightarrow -\omega_a$ and  $\Delta E_\alpha \rightarrow \omega_\alpha$. This leads finally to
\begin{equation}\label{SIFF-derivation}
    \frac{\mel{A(\vb{p}+\vb{q}_a)}{J_{\mu_a}}{A(\vb{p})}}{-\Delta E_a+ \iu 0}\rightarrow 
    \frac{(Z_Ae)v_{\mu_a}F_A(\vb{q}_a^2)}{\Delta \omega_a+ \iu 0} = \frac{(Z_A e) F_A(\vb{q}_a^2)  v_\mu }{v\cdot q_a + \iu 0} ~,
\end{equation}
where $v_\mu=(1,0,0,0)$ is the four velocity of the heavy composite in its rest frame. This is precisely what one would obtain using a Wilson line or heavy particle Feynman rules, but with a form factor included at every vertex.
 
We now have simplifying expressions for generic correlators in the limit $R^{-1} \sim q \ll M$. This simplifying limit of the correlators naturally appears when one considers loop-momenta regions where $q_i \sim R^{-1}\ll M$ \cite{Beneke:1997zp,Jantzen:2011nz}; we will call this the soft region. Hard regions $q \gg R^{-1}$ will factorize from soft regions. We therefore expect the soft region, where the hadronic correlators have a simplified form and  where $Z$-enhanced contributions exist,  to be described by a corresponding EFT which we now describe.

\section{Construction of the effective theory \label{NRQED} }
In the previous section we have seen that an analysis of correlation functions of heavy particles naturally yields Feynman rules with  eikonal propagators and  structure-dependent vertex functions.  It is then natural to ask whether the soft region presented above has its own EFT. We claim that it does, and  that the degrees of freedom are the Wilson lines traced out by a finite charge density moving along an eikonalized world line. 

\subsection{Construction from NRQED} 
In the limit where $q \ll 1/R$ we must be able to match onto NRQED \cite{Caswell:1985ui,Manohar:1997qy,Hill:2012rh,Paz:2015uga}. For large composite objects, the charge radius and  other higher multipoles will be ``unaturally large'' in the power counting of NRQED because  $1/M \ll R$. Counting powers of $R$ and  $M$ independently then allows one to consistently account for higher derivative operators that are enhanced by the large nuclear radius while still performing an expansion in $1/M$. Beyond the leading kinetic terms, the NRQED Lagrangian in the one-particle sector can be constructed out of $\vb{E}$, $\vb{B}$, and  the non-relativistic field $\psi_v$ which carries the four-velocity label $v$.

Importantly $\vb{B}$ is odd with respect to parity and  must be accompanied by a non-trivial spin structure, e.g.\ $\vb*{\sigma}\cdot \vb{B}$. By contrast $\vb{E}$ is even with respect to parity and  this therefore allows operators of the form $\vb*{\partial}\cdot \vb{E}$ and  more generally $(\vb*{\partial}^2)^n \vb*{\partial}\cdot \vb{E}$. The Wilson coefficients of these operators may be determined by a matching calculation that equates $\mel{\vb{p}+\vb{q}}{J_\mu}{\vb{p}}$ in both the full theory and  NRQED (see e.g.\ \cite{Hill:2012rh} for details). As we have discussed in \cref{static-form-factors}, if we neglect terms that are suppressed by powers of $|\vb{q}|/M$ we only need the charge form factor $F_1(q^2)$. In this limit we may write the  Lagrangian as
\begin{equation}
     \label{simple-NRQED}
    \mathcal{L}_{\rm EFT} = \psi_v^\dagger (\iu v\cdot D) \psi_v
    + \sum_{n=1}^\infty C_{\vb*{\partial} \cdot \vb{E} }^{(n)} \psi_v^\dagger \psi_v  ~\qty(\vb*{\partial}^2)^{n-1} \vb*{\partial} \cdot \vb{E}~ +O(|\vb{q}|/M)~. 
\end{equation}
Matching in the one particle sector then dictates 
\begin{equation}
  C_{\vb*{\partial}\cdot \vb{E}}^{(n)} = \frac{\overline{F}^{(n)} }{n!} ~ + O(Z|\vb{q}|/M)~.
\end{equation}
These are the unique set of terms that contribute to $\mathcal{L}_{\rm NRQED}$ that are {\it i)} coherently enhanced, {\it ii)} survive in the $M\rightarrow \infty$ limit.

Notice that in \cref{simple-NRQED} the minimal coupling term, $\iu v\cdot D$, is unique in that it contributes to the matching of $F(q^2)$, and contains the scalar potential $v\cdot A$. At higher orders in $1/M$ one can always remove factors of $v\cdot A$ with the equations of motion. We may re-sum the full set of $R$-enhanced Wilson coefficients\footnote{See \cite{Aoude:2020ygw,Haddad:2020que} for a similar discussion but where the spin, rather than the radius, of the object is counted as large.}, using the fact that $\sum_{n=1}^\infty \frac{\overline{F}^{(n)} }{n!} (q^2)^{n-1} = f(q^2)$ where $f(q^2)=(F(q^2)-1)/q^2$ we find 
\begin{equation}
  \label{resum-NRQED}
  \mathcal{L}_{\rm EFT} = \psi^\dagger (\iu v\cdot D) \psi
  + \psi_v^\dagger \psi_v^{~} f(\vb*{\partial}^2) \vb*{\partial}\cdot \vb{E} + O(|\vb{q}|/M)~.
\end{equation}
The Feynman rule for a 1-photon vertex generated by \cref{resum-NRQED} agrees order-by-order in the derivative expansion with \cref{simple-NRQED} up to corrections of $O(|\vb{q}|/M)$. It is given explicitly by
\begin{equation}  
    \begin{split}
  \label{resum-feynmanrule}
 \raisebox{-10pt}{\begin{fmffile}{1coul-a}
    \begin{fmfgraph*}(65,35) 
    \fmfbottom{i1,d1,o1}
    \fmftop{i2,d2,o2}
    \fmf{double}{i1,b1,o1}
    \fmf{phantom}{o2,t1,i2}
    \fmf{photon,tension=0}{b1,t1}
    \fmfblob{0.2w}{b1}
\end{fmfgraph*}
\end{fmffile}}
    &= ~~\iu  (Ze)v_\mu  + \iu (Ze) \frac{F(\vb{q}^2)-1}{-\vb{q}^2}\qty[ v_\mu q^2 - q_\mu v\cdot q]\\
    &=~~\iu (Ze) v_\mu + \iu (Ze) \frac{F(\vb{q}^2)-1}{q_\perp^2}\qty[ v_\mu q_\perp^2 - q_{\perp\mu} v\cdot q]~,
    \end{split}
\end{equation}

\vspace{10pt}

\noindent where $\vb{q}_{\perp\mu}=q_\mu-(v\cdot q)v_\mu=(0,\vb{q})$. 
Notice that if this expression is multiplied by $\delta(v\cdot q)$ then the Feynman rule reduces to 
\begin{equation}
    \iu v_\mu (Ze)  + \iu (Ze) \frac{F(\vb{q}^2)-1}{q_\perp^2}\qty[ v_\mu q^2 - q_\mu v\cdot q] \rightarrow  \iu v_\mu (Ze) F(\vb{q}^2) \qq{for} v\cdot q=0~,
\end{equation}
as one would naively expect for a static background field. 
 
Being constructed from the gauge invariant quantity $\vb*{\partial}\cdot \vb{E}$, current conservation is automatically enforced for terms proportional to $f(\vb{q}^2)$. The first term in \cref{resum-feynmanrule} comes from the minimal coupling from the kinetic term $\psi^\dagger_v(v\cdot D) \psi_v$. When included in a full class of diagrams, including any light/dynamical charged particles present in the reaction, the resultant sum will satisfy current conservation. A similar resummation of terms enhanced by $(qR)^n$ at subleading powers in $1/M$ (e.g.\ for the dipole density distribution) may be performed at higher power, although we do not pursue this here. The analysis is straightforward involving the identification of terms from a given form factor which contribute order-by-order in the NRQED expansion.

The minimal coupling part of the Lagrangian will satisfy Ward identities when combined with other parts of the amplitude. The structure dependent piece is manifestly gauge invariant vertex-by-vertex. Taken together this then ensures that the relevant Ward identities will be satisfied and  that the inclusion of form factors inside loops will produce gauge invariant amplitudes. See \cref{Gauge_invariance} for a more detailed argument. 

\subsection{Wilson lines and  decoupling transformations}
Having now understood that the EFT of massive extended objects is a natural generalization of NRQED with modified power counting, we have arrived at Feynman rules that are very similar to those for point-like Wilson lines. In what follows, we will construct the operator definition of a Wilson line for a finite charge distribution that reproduces the Feynman rules derived above. The definition is non-local and simplifies dramatically in  Coulomb gauge. 

Using coordinates such that $x_\mu = (\vb{x},v\cdot x)$, the QED Wilson line, for a point-particle of charge $Z$, is given by \cite{Wilson:1974sk,Becher:2014oda,Grozin:2022umo}
\begin{equation}
    \begin{split}
    \widehat{W}_v(x)&= \exp\qty [\iu Z e \int_0^\infty \dd s~~v\cdot \widehat{A}(\vb{x}, v\cdot x+ s ) ]~, \\
    &=\exp\qty [\iu Z e  \int_0^\infty \dd s~\int \dd^3y ~ \delta^{(3)}(\vb{x}-\vb{y})v\cdot \widehat{A}(\vb{y}, v\cdot x + s) ] . 
    \end{split}
\end{equation} 
In the second equality we have suggestively re-written the standard definition in terms of a three dimensional point-like charge distribution. One might guess a natural generalization for an extended charge distribution is to smear the vector potential with a charge distribution, 
\begin{equation}
    \begin{split}
      \widehat{W}_{v,\rho}(x)
    &=\exp\qty [\iu Z e  \int_0^\infty \dd s~\int \dd^3y ~ \rho(\vb{y}-\vb{x})v\cdot \widehat{A}( \vb{y}, v\cdot x + s )] \qq{(naive).}
    \end{split}
\end{equation} 
however this turns out not to be gauge invariant. The correct expression in general gauge is given by 
\begin{equation}
     \label{ex-WL-def}
    \begin{split}
    \widehat{W}_{v,\rho}(x)
    &=\widehat{W}_{v}(x) \e^{\iu Ze \chi }~\qq{(general gauge).}  
    \end{split}
\end{equation} 
where $\chi$ is given by, 
\begin{equation}
     \label{chi-WL-def}
    \begin{split}
     \chi&=\int \dd^3 y ~\qty(\rho(\vb{x}-\vb{y})-\delta^{(3)}(\vb{x}-\vb{y}) ) \frac{1}{\vb*{\partial}^2}  [\vb*{\partial}^2 (v\cdot A) - (v\cdot \partial)\vb*{\partial}\cdot \vb*{A} ](\vb{y},v\cdot x + v_\mu s)~,\\
     &= \int \dd^3 z \int \dd^3 y ~\qty(\rho(\vb{x}-\vb{y})-\delta^{(3)}(\vb{x}-\vb{y}) ) \frac{1}{|\vb{y}-\vb{z}|} \\
     &\hspace{0.4\linewidth}[\vb*{\partial}^2 (v\cdot A) - (v\cdot \partial)\vb*{\partial}\cdot \vb*{A} ](\vb{z},v\cdot x + s)~.
     \end{split}
\end{equation}
This operator reproduces the Feynman rule \cref{resum-feynmanrule} in arbitrary gauge.   Setting $\vb*{\partial} \cdot \vb{A}=0$ we arrive at the expression in Coulomb gauge
\begin{equation}
    \begin{split}
      \widehat{W}_{v,\rho}(x)
    &=\exp\qty [\iu Z e  \int_0^\infty \dd s~\int \dd^3y ~ \rho(\vb{y}-\vb{x})v\cdot \widehat{A}( \vb{y} , v\cdot x + s)] ~\text{(Coul. gauge)~,} 
    \end{split}
\end{equation} 
which agrees with the naive expectation.  It is interesting to see how this works in a diagramatic language where Coulomb gauge amounts to the following choice for the photon propagator,
\begin{equation}
    D^{\mu\nu}=  -\frac{\iu}{q_\perp^2} v^\mu v^\nu  - \frac{\iu}{q^2+\iu 0}\qty[(-g^{\mu\nu} + v^\mu v^\nu) +\frac{q_\perp^\mu q_\perp^\nu}{q_\perp^2} ]~,
\end{equation}
such that $D^{00}=\iu/\vb{q}^2$ for $v_\mu=(1,0,0,0)$. Using the Feynman rule for the vertex in \cref{resum-feynmanrule}  we arrive at, 
\begin{equation}
    \Gamma_\mu D^{\mu\nu} = J_0 D^{0\nu} - J_i D^{i\nu}=
    J_0 D^{00} + J_i D^{ij}  = \qty(\iu (Ze)F(\vb{q}^2)\frac{\iu}{\vb{q}^2}   ,  0 )
\end{equation} 
where we have used $q_\perp^i (\delta^{ij} - q_\perp^i q_\perp^j/\vb{q}^2) =0$. One then sees that the extended Wilson line vertex, when contracted with a Coulomb-gauge propagator reduces to the naive expectation of a static Coulomb field with a charge form factor.

One can check that the above Wilson line reproduces the Feynman rules derived from the heavy particle Lagrangian above. 
It would be interesting to develop factorization theorems in terms of these extended Wilson lines rather than their point-like counterparts, however we leave this for future work. 

\subsection{Emergence of a Coulomb field in the static limit \label{Coulomb-static} }
It is well known that ladder graphs resum in QED to produce the Coulomb potential. This was first analyzed in the context of a relativistic Dirac fermion \cite{Brodsky:SLAC1010,Dittrich:1970vv,Neghabian:1983vm}. Weinberg demonstrated that this result was universal, independent of the spin of the particle, applying for any system when momenta transfer are sufficiently low that one is in the point-like limit \cite{Weinberg:1995mt}. All of Refs.\ \cite{Brodsky:SLAC1010,Dittrich:1970vv,Neghabian:1983vm,Weinberg:1995mt} restrict themselves to the following setup 
\begin{itemize}
    \item Elastic scattering. 
    \item Charge conservation in the heavy sector. 
    \item The point-like limit (i.e.\ $q_i R\ll 1$ for all photon momenta in the graph). 
\end{itemize}
The analysis presented in this work dispense with all of these assumptions and demonstrates that the electrostatic field of an extended charge density distribution emerges in the static limit independent of the above assumptions provided the source is effectively static i.e., that $|q|\ll M$.  

Let us denote the vertex function from the EFT by $\Gamma_\mu(q_i)$. It will be convenient to phrase out discussion in terms of hadronic correlators, $G_{\{\mu_i\}}(\{q_j\})$,  but now computed directly in the EFT.\footnote{This is equivalent to including only elastic excitations of the nucleus in the full theory. It captures the leading-in-$Z$ effects.} We have, 
\begin{equation}
    \label{G-el-AequalB}
    \begin{split}
     G_{\{\mu_i\}}(\{q_j\})=\sum_{\rm perms}\prod_{a}^{N_L}\frac{ Z \Gamma_{\mu_a}(q_a)}{v\cdot q + \iu 0}   \prod_{\alpha}^{N_R} \frac{ Z \Gamma_{\mu_\alpha}(q_\alpha)}{-v'\cdot q_\alpha + \iu 0} ~\times \mathcal{J}_{AB} ~, 
     \end{split}
\end{equation}
where $\mathcal{J}_{AB}$ is an external operator we have inserted in the theory. Notice that the sum over eikonal propagators in \cref{G-el-AequalB} is equivalent to a product of delta functions which set every factor of $v \cdot q_i=0$ \cite{Weinberg:1995mt},
\begin{equation}
    \label{simplified-G-el}
     G_{\{\mu_i\}}(\{q_j\}) = \mathcal{J}_{AB}\qty[\prod_i (2\pi \iu) \delta(v\cdot q_i) Z_A \Gamma_{\mu_i}(q_i)]  = \mathcal{J}_{AB}\qty[\prod_i (2\pi \iu) \delta(v\cdot q_i) Z_A v_{\mu_i} F(\vb{q}^2_i)] ~. 
\end{equation}
where we have used $\Gamma_\mu(q) =v_{\mu} F(\vb{q}^2)$ for $v\cdot q=0$. Theses are the Feynman rules for a background Coulomb field sourced by a static charge density distribution, $\rho(\vb{x})$, whose Fourier transform gives $F(\vb{q}_i^2)$.

In deriving \cref{simplified-G-el} one must properly account for infinitesimal imaginary parts in the denominator i.e.\ $1/(v\cdot q_\alpha +\iu 0)$ vs $1/(-v\cdot q_a+ \iu 0)$. If $\mathcal{J}_{AB}$  is a charged current then the leading-$Z$ behavior is still Coulomb in nature, and new generalized eikonal identities 
can be used to simplify sub-leading contributions \cite{Plestid:2024eib}.

\section{Applications to $(e,e')$ scattering \label{e-e'} } 
In this section we will consider electron nucleus scattering at large momentum transfer, 
 \begin{equation}
    e(\vb{k}) ~ + ~A(\vb{p}) ~\rightarrow ~ \ell^-(\vb{k}') ~+~ B(\vb{p}') ~, 
 \end{equation}
in the limit $R^{-1}\ll |Q| \ll M$ where $Q^2=(p-p')^2$. 
 In this case the hard current is the electromagnetic current itself $\mathcal{J}_\nu=J_\nu$. At tree level the amplitude is given by
\begin{equation}
    \label{0th-e-eprime}
    \iu \mathcal{M}^{(0)} = \iu e^2\frac{1}{Q^2} \bar{u}(k)\gamma^\nu u(k) \mel{B(\vb{p}+\vb{Q})}{\mathcal{J}_\nu}{A(\vb{p})}~.
\end{equation}
At 1-loop, ladder diagrams appear coupling the lepton to both the initial and  final state nucleus. 
\begin{equation}
    \label{one-loop-e-eprime}
    \begin{split}
    \iu \mathcal{M}^{(1)} &= \iu e^4\int \frac{\dd^4 q_1}{(2\pi)^4} ~\frac{1}{q_1^2} \frac{1}{(q_1-Q)^2} \frac{\bar{u}(k')\gamma^{\nu}(\slashed{k}+\slashed{q}_1+m)\gamma^{\mu_1} u(k)}{2k\cdot q_1 + q_1^2} G_{\mu_1;\nu}(q_1)\\
    &\hspace{0.1\linewidth}+
    \iu e^4\int \frac{\dd^4 q_1}{(2\pi)^4} ~\frac{1}{q_1^2} \frac{1}{(q_1-Q)^2} \frac{\bar{u}(k')\gamma^{\nu}(\slashed{k}'-\slashed{q}_1+m)\gamma^{\mu_1} u(k)}{-2k'\cdot q_1 + q_1^2} G_{\mu_1;\nu}(q_1)~.
    \end{split}
\end{equation}
At $n^{\rm th}$ order in perturbation theory we will encounter $G_{\mu_1,... \mu_n;\nu}(q_1...q_n)$.

We introduce the power counting parameter $\lambda = (QR)^{-1}$. Then, using the method of regions, we may split the integral into a region where $q\sim 1$ and  $q\sim  \lambda$ respectively. We will refer to these as the soft and  hard regions respectively. Coherent enhancements occur exclusively in the soft region where the theory is described by the EFT outlined above. The matrix element of this EFT, $\mathcal{M}_S^{(1)}$ may be obtained from \cref{one-loop-e-eprime} by counting $q\sim \lambda$ and  keeping leading order terms in $\lambda$. 

\subsection{Eikonal approximation} 
In the limit $QR\gg 1$, it is well known that eikonal approximation describes $(e,e')$ data well \cite{Yennie:1965zz,Aste:2004kh,Aste:2005wc,Tjon:2006qe}.  This corresponds to the matrix element evaluated in the soft region. Choosing $v_\mu=(1,0,0,0)$, setting the electron mass to zero, and  performing the integral over $q_0$ via a contour integral, using $\gamma_0 \gamma_i \gamma_0 = -\gamma_i$ and Dirac algebra on the external spinors  we find
\begin{equation}
    \label{eikonal} 
     \mathcal{M}_S^{(1)} = \iu   \mathcal{M}^{(0)} \times Ze^2\int \frac{\dd^3 q}{(2\pi)^3}~\frac{F(\vb{q}^2)}{\vb{q}^2-\iu 0}  \qty[\frac{2E(k)}{2\vb{k}\cdot \vb{q}-\iu 0} +\frac{2E(k')}{-2\vb{k}\cdot \vb{q}-\iu 0}]~. 
\end{equation}  
This can be recognized as the Feynman rule for an ultra-relativistic eikonalized field (as appears in soft collinear effective theory) interacting with a static charge distribution, with charge form factor $F(\vb{Q}^2)$. It reproduces the $O(Z\alpha)$ term in the standard eikonal phase.  The result relies essentially on the scale introduced by $F(\vb{q}^2)$ without which the integral would be scaleless and  vanish in dimensional regularization. Nuclear structure is therefore essential to the eikonal  approximation. 

\subsection{Structure mismatch} 
For inelastic scattering $A\rightarrow B$, we may have $F_A(Q^2) \neq F_B(Q^2)$. There will be ``mismatch'' between photon exchange with $A$ and  $B$ even for equal charges. The results presented above  allows one to calculate this effect systematically in perturbation theory.  As an illustration we work at one loop in the static limit. Expanding around the high-energy limit we introduce a light-like reference vector $n_\mu =(1,0,0,1)$ writing $k_\mu = E_e n_\mu$.   Defining $\delta F= F_A-F_B$ we have   
\begin{equation}
  \int \frac{\dd^4 L}{(2\pi)^4} ~\frac{\delta F(\vb{L}^2) }{\vb{L}^2}\qty[ v_\mu L^2 - L_\mu v \cdot L]  \frac{1}{v\cdot L+ \iu 0} \frac{n^\mu}{n\cdot L + \iu 0}
  \frac{1}{L^2 + \iu 0}~.
\end{equation}
We perform integrals over $q_0$ using contour integration  since $F(\vb{q}^2)$ does not depend on $q_0$. The result is that the contributions at one-loop order proportional to $\delta f$ can be written as 
\begin{equation}
  \int \frac{\dd^4 L}{(2\pi)^4} ~\frac{\delta F(\vb{L}^2)}{\vb{L}^2} \frac{1}{L^2 + \iu 0}
  =   \frac{\iu }{2}\int \frac{\dd^3 L}{(2\pi)^3} ~\delta F(\vb{L}^2)\qty[\frac{1}{\vb{L}^2}]^{3/2}~.
\end{equation}
Taking $F_{A,B}(\vb{L}^2)= \Lambda_{A,B}^2/(\vb{L}^2 + \Lambda_{A,B}^2)$ gives for the difference of form factors,  
\begin{equation}
    \delta F(\vb{L}^2)= \frac{\vb{L}^2(\Lambda_A^2-\Lambda_B^2)}{(\vb{L}^2+\Lambda_A^2)(\vb{L^2}+\Lambda_B^2)}~,
\end{equation}
which gives 
\begin{equation}
  \frac{\iu }{2}\int \frac{\dd^3 L}{(2\pi)^3} ~\delta F(\vb{L}^2)\frac{1}{|\vb{L}|^3} = \frac{\iu}{4\pi^2} \log\qty(\frac{\Lambda_A}{\Lambda_B})~,
\end{equation}
for the structure mismatch introduced above. At this order the resulting correction only affects the imaginary part of the amplitude which then enters at $O(\alpha^2)$ in the cross section. 

\vfill
\pagebreak

\section{Conclusions \label{conclusions}}

We have constructed an EFT appropriate for the interaction of heavy, extended, composite objects with electromagnetic probes. Our work provides a generalization of of Wilson lines for heavy but spatially extended charge distributions. The fact that such a generalization exists should not be surprising since a spatial Fourier transform of a charge form factor becomes well defined in the infinite mass limit.  The theory supplies a systematic framework to investigate structure dependent corrections for heavy particles at higher orders in a loop expansion.

The EFT allows us to straddle the high and  low momentum limits, where $pR\gg1$ and  $pR\ll1$ respectively, within a unified framework.  By including the extended Wilson lines in the ``full theory'' we may identify modes using the method of regions for different kinematic configurations. This allows us to construct the relevant EFT for the problem at hand, and  does not artificially restrict our analysis to the point-like limit. Indeed for the high energy limit this is essential, since the eikonal expansion for potential  scattering is ill-defined in the point-like limit. For low-momentum probes (e.g.\ in nuclear beta decays) the structure dependent contribution modifies short-distance Wilson coefficients.

An extension to sub-leading power would be interesting. We expect a similar identification of ``radius enhanced'' operators in NRQED can be carried out at sub-leading power (e.g.\ for the dipole density), and an extension of the same ideas to the multi-photon sector e.g.\ contributions from nuclear polarizabilities, may also be of interest. Nuclear excited states may be straightforwardly included by allowing for additional heavy particles with an explicit mass splitting in the theory.  It would be interesting to apply the concepts presented here to more general settings. In particular, gravity where large coherent enhancements for heavy extended objects are ubiquitous (e.g.\ for black holes and neutron stars), and structure dependent QED corrections in heavy meson decays where the radius $R\sim 1/\Lambda_{\rm QCD}$ is parametrically separated from the mass $M\sim m_b ~{\rm or}~ m_c$. 

\vfill
\pagebreak

\subsection*{Acknowledgments}
I thank Richard J.\ Hill \& Oleksandr Tomalak for useful discussions during collaboration on related work, and Michele Papucci, Julio Para-Martinez, Ira Rothstein, Florian Herren, and Andreas Helset for useful feedback. Part of this work was completed at the Kavli Institute for Theoretical Physics during the program ``Neutrinos as a Portal to New Physics and  Astrophysics'' and  I thank KITP for their hospitality. KITP is  supported in part by the National Science Foundation under Grant No.\ NSF PHY-1748958. This work benefited from feedback and  discussions during the workshop ``Power expansions on the lightcone'' at the Mainz Institute for Theoretical Physics and  I would like to thank them for their hospitality. This work is supported by the Neutrino Theory Network under Award Number DEAC02-07CH11359, the U.S. Department of Energy, Office of Science, Office of High Energy Physics, under Award Number DE-SC0011632, and by the Walter Burke Institute for Theoretical Physics.

\vfill
\pagebreak

\appendix 

\section{Gauge invariance of background field terms \label{Gauge_invariance} } 
Correlators which are proportional $G_{\mu}(q_i) \propto v_\mu \delta(v\cdot q_i)$ automatically satisfy current conservation, i.e.\ $q_i^\mu G_\mu (q_i)=0$. One then expects that relevant Ward identities being satisified, the insertion of blobs involving $G_\mu$ inside of Feynman diagrams will generate gauge invariant amplitudes. In a general gauge, however, the photon propagator can be expressed as 
\begin{equation}
    D_{\mu \nu}(q) = \iu \qty[ -g_{\mu\nu} + q_\mu c_\nu(q) + c_\mu(q) q_\nu]/q^2~.
\end{equation}
with $c_\nu(q)$ not necessarily proportional to $q_\nu$. 

For gauge invariance we must therefore require that both $(v\cdot q) \delta(v\cdot q)=0$ and   $(v\cdot c) \delta(v\cdot q)$ vanish. At first glance, this seems not to hold since $c_\nu(q)$ is a generic function of $q$, however closer inspection reveals that if  $ (c\cdot q)\delta(v\cdot q)=0$ in one gauge then it must also hold in arbitrary gauge. This condition can be explicitly checked in Coulomb gauge, axial gauge, and  arbitrary $R_\xi$ gauge and  one finds that indeed $(c\cdot q)\rightarrow 0$ when $v\cdot q \rightarrow 0$.

To see why this must be the case in an arbitrary gauge consider $A_\mu \rightarrow A_\mu + \partial_\mu \Lambda$ (at the operator level we take $\Lambda$ to be linear in the creation and  annihilation operators of the photon field). Consider the shift in the propagator
\begin{equation}
    D_{\mu \nu} = \int \dd^4 x ~\e^{-\iu q\cdot x} \mel{0}{T\{ A_\mu (x) A_\nu(0) \}}{0}~.
\end{equation}
Let us consider specifically $v^\mu D_{\mu \nu}$. Terms with $v\cdot \partial \Lambda$ will be proportional to $v\cdot q$ and  will vanish when multiplied by $\delta(v\cdot q)$. We then only have to worry about the term $\mel{0}{T\{ v\cdot A(x)  \partial_\nu\Lambda(0) \}}{0}$. We may reach an arbitrary gauge via a gauge transformation from any of axial gauge, Coulomb gauge, or $R_\xi$ gauge, and  since  $v\cdot A\propto v\cdot q$ in all of these gauges, the contribution to the time-like component of the photon propagator, $v^\mu D_{\mu\nu}$, will vanish when multiplied by $\delta(v\cdot q)$. This then demonstrates that 
\begin{equation}
    \delta(v\cdot q) v^\mu D_{\mu \nu}(q) = -\iu v_\nu \delta(v\cdot q)/q^2 = -\iu v_\nu \delta(v\cdot q)/(-\vb{q}^2)~,
\end{equation}
holds in an arbitrary gauge as desired.   

\vfill
\pagebreak
\bibliography{Ex_WilsonLine.bib}
\end{document}